\documentclass[a4paper,11pt]{article}
\usepackage[utf8]{inputenc}
\usepackage[spanish]{babel}
\usepackage{amsmath}
\usepackage{amsfonts}
\usepackage{amssymb}
\usepackage{appendix}  
\usepackage{graphicx}
\usepackage{wrapfig}
\usepackage{caption, subcaption}
\usepackage[left=2.5cm,right=2.5cm,top=2.5cm,bottom=2.5cm]{geometry}
\usepackage{xcolor}
\usepackage{algorithm}
\usepackage{algpseudocode}

\usepackage{listings} 

\usepackage{xcolor} 

\lstdefinestyle{mystyle}{
    commentstyle=\color{gray},       
    keywordstyle=\color{blue},       
    numberstyle=\tiny\color{gray},   
    stringstyle=\color{red},         
    basicstyle=\small\ttfamily,      
    breakatwhitespace=false,         
    breaklines=true,                 
    captionpos=b,                    
    keepspaces=true,                 
    numbers=left,                    
    numbersep=5pt,                   
    showspaces=false,                
    showstringspaces=false,          
    showtabs=false,                  
    tabsize=2,                       
    frame=single,                    
    lineskip=0.5pt                   
}
\usepackage{float} 

\title{Desarrollo de competencias STEM mediante la programación de modelos de auto-organización}
\author{Dr. Hernández Rodríguez, Matías Ezequiel}
\date{\today}

\begin{document}

\maketitle

\begin{abstract}
Este artículo presenta una propuesta educativa basada en la implementación computacional de un modelo de interacción de partículas de diferentes tipos, como herramienta para el desarrollo de competencias STEM (Ciencia, Tecnología, Ingeniería y Matemáticas). Se detalla la formulación matemática del modelo, las ecuaciones de interacción entre partículas, la matriz de interacción que gobierna el comportamiento colectivo, y un pseudocódigo completo del algoritmo de simulación. Se discute cómo este proyecto interdisciplinario permite a los estudiantes aplicar y conectar conocimientos de física, matemáticas, programación y pensamiento sistémico, fomentando habilidades como el pensamiento computacional, la modelización matemática y la comprensión de sistemas complejos.
\end{abstract}

\selectlanguage{english}
\renewcommand{\abstractname}{Abstract}  

\begin{abstract}
This article presents an educational proposal based on the computational implementation of a model of interaction between particles of different types, as a tool for the development of STEM (Science, Technology, Engineering, and Mathematics) skills. The mathematical formulation of the model, the equations governing particle interaction, the interaction matrix that governs collective behavior, and a complete pseudocode of the simulation algorithm are detailed. The article discusses how this interdisciplinary project enables students to apply and connect knowledge from physics, mathematics, programming, and systems thinking, fostering skills such as computational thinking, mathematical modeling, and understanding of complex systems.
\end{abstract}

\selectlanguage{spanish}

\section{Introducción}

El enfoque educativo STEM, que busca integrar la enseñanza de Ciencia, Tecnología, Ingeniería y Matemáticas, tiene como objetivo principal preparar a los estudiantes para enfrentar y resolver los problemas complejos que caracterizan el mundo real. Este enfoque interdisciplinario no solo proporciona una formación técnica sólida, sino que también fomenta el pensamiento crítico, la creatividad y la capacidad de resolver problemas de manera innovadora. A través de la integración de estas disciplinas, los estudiantes adquieren un conocimiento más profundo y una mayor comprensión de los principios fundamentales que rigen tanto los fenómenos naturales como los creados por el ser humano. La enseñanza de estos principios prepara a los estudiantes para abordar desafíos multifacéticos que requieren un enfoque global y flexible \cite{Aguirre2019, Bautista2021, Bravo2018, Santillan2020, Silva2023}.

Uno de los aspectos más interesantes de este enfoque interdisciplinario es la conexión entre los fenómenos de auto-organización y emergencia, los cuales representan una excelente oportunidad para la educación STEM. Estos fenómenos, que surgen naturalmente en sistemas complejos, pueden ser utilizados para ilustrar cómo principios de diversas disciplinas pueden converger para ofrecer soluciones más completas y efectivas a problemas complejos.

La auto-organización hace referencia a la capacidad de un sistema de desarrollar estructuras ordenadas y complejas a partir de condiciones iniciales desordenadas, sin necesidad de un control externo centralizado. Este fenómeno puede observarse en diversos sistemas naturales, como los ecosistemas, las formaciones de las colonias de insectos, o incluso en el comportamiento colectivo de los seres humanos \cite{Camazine2003}. En un contexto educativo, la auto-organización puede ser utilizada para ayudar a los estudiantes a comprender cómo los sistemas dinámicos pueden evolucionar y adaptarse de manera autónoma, lo que resulta útil en diversas áreas, como la ingeniería, la biología y la computación.

Por otro lado, la emergencia describe el fenómeno mediante el cual las interacciones locales entre los componentes individuales de un sistema pueden dar lugar a comportamientos globales complejos que no son evidentes a partir de las reglas subyacentes. Este concepto, ampliamente estudiado en las ciencias físicas, biológicas y sociales, muestra cómo patrones complejos pueden surgir sin que haya una planificación explícita o control desde una entidad central \cite{Holland1998}. En el ámbito educativo, la emergencia puede utilizarse para ilustrar cómo los estudiantes pueden analizar y modelar fenómenos complejos en los que pequeños cambios locales tienen un impacto significativo en el comportamiento global del sistema.

Ambos fenómenos, auto-organización y emergencia, no solo enriquecen el enfoque STEM, sino que también ofrecen una excelente manera de conectar conceptos de diferentes disciplinas, permitiendo que los estudiantes comprendan cómo los principios fundamentales de la ciencia, la tecnología, la ingeniería y las matemáticas se interrelacionan de manera profunda y aplicable a situaciones reales.

En este artículo, estudiamos cómo la implementación de un modelo inspirado en el trabajo \textit{How a life-like system emerges from a simple particle motion law} \cite{Schmickl2017} puede utilizarse como proyecto educativo para desarrollar estas competencias. 

El artículo está organizado de la siguiente manera. En la sección 2, se expone la formulación matemática del modelo de interacción de partículas propuesto. La sección 3 introduce el concepto de matriz de interacción. En la sección 4, se aborda la discretización necesaria para la simulación numérica del modelo, un aspecto fundamental para su implementación computacional. La sección 5 presenta el pseudocódigo correspondiente al algoritmo asociado al modelo. En la sección 6, se analizan las competencias STEM promovidas por la implementación de este modelo en un contexto educativo. Las secciones 7 y 8 están dedicadas, respectivamente, a la propuesta de actividades didácticas y a la enumeración de las evaluaciones de competencias derivadas del modelo. Finalmente, en la sección 9, se presentan las conclusiones del estudio y las líneas de investigación futura del proyecto.

\section{Formulación matemática del modelo}

El modelo consiste en $N$ partículas que se mueven en un espacio bidimensional con condiciones de frontera periódicas (toroidal). Cada partícula $i$ tiene una posición $\vec{r}_i = (x_i, y_i)$, una velocidad $\vec{v}_i = (v_{x,i}, v_{y,i})$, y pertenece a uno de los $T$ tipos posibles.

\subsection{Ecuaciones de movimiento}

La dinámica de cada partícula está gobernada por las siguientes ecuaciones diferenciales:

\begin{equation}
\frac{d\vec{r}_i}{dt} = \vec{v}_i
\end{equation}

\begin{equation}
\frac{d\vec{v}_i}{dt} = \alpha\sum_{j \neq i} \vec{F}_{ij} - \gamma\vec{v}_i,
\end{equation}
donde $\alpha$ es un parámetro que controla la magnitud de las fuerzas, $\vec{F}_{ij}$ es la fuerza que ejerce la partícula $j$ sobre la partícula $i$, y $\gamma$ es un coeficiente de amortiguamiento que previene que el sistema se vuelva inestable.

\subsection{Función de fuerza}

La fuerza entre dos partículas depende de su distancia y de sus tipos. Sean $\tau_i$ y $\tau_j$ los tipos de las partículas $i$ y $j$, respectivamente, y sea $r_{ij} = |\vec{r}_j - \vec{r}_i|$ la distancia entre ellas. La magnitud de la fuerza se define como:

\begin{equation}
F(r_{ij}, \tau_i, \tau_j) = 
\begin{cases}
\frac{r_{ij}}{\beta r_{max}} - 1, & \text{si } r_{ij} < \beta r_{max} \\
G_{\tau_i, \tau_j} \left(1 - \left|2\left(\frac{r_{ij}}{r_{max}} - 0.5\right)\right|\right), & \text{si } \beta r_{max} \leq r_{ij} < r_{max} \\
0, & \text{si } r_{ij} \geq r_{max},
\end{cases}
\end{equation}
donde $r_{max}$ es la distancia máxima de interacción, $\beta$ es un parámetro que determina el radio de repulsión relativo a $r_{max}$, y $G_{\tau_i, \tau_j}$ es el elemento de la matriz de interacción que define la fuerza entre partículas de tipos $\tau_i$ y $\tau_j$.

La dirección de la fuerza es paralela al vector que une las partículas:

\begin{equation}
\vec{F}_{ij} = F(r_{ij}, \tau_i, \tau_j) \frac{\vec{r}_j - \vec{r}_i}{r_{ij}}
\end{equation}

\subsection{Interpretación de la función de fuerza}

La función de fuerza tiene tres regímenes que los estudiantes pueden analizar para comprender conceptos físicos fundamentales:

\begin{enumerate}
    \item \textbf{Repulsión a corta distancia:} Cuando $r_{ij} < \beta r_{max}$, la fuerza es repulsiva independientemente de los tipos de partículas. Esto permite explorar principios como el de exclusión o las interacciones electrostáticas.
    
    \item \textbf{Interacción específica de tipo:} En el rango $\beta r_{max} \leq r_{ij} < r_{max}$, la fuerza puede ser atractiva o repulsiva dependiendo del valor $G_{\tau_i, \tau_j}$ en la matriz de interacción. Los estudiantes pueden relacionar esto con fuerzas selectivas en química o biología.
    
    \item \textbf{Sin interacción:} Más allá de $r_{max}$, las partículas no interactúan, lo que introduce conceptos de optimización computacional y localidad de interacciones.
\end{enumerate}

\section{Matriz de interacción}

La matriz de interacción $G$ es una matriz de dimensiones $T \times T$ donde cada elemento $G_{ij}$ define la fuerza de interacción entre partículas de tipo $i$ y tipo $j$. Este componente ofrece una oportunidad valiosa para que los estudiantes exploren conceptos de álgebra lineal y representación matricial:

\begin{itemize}
    \item $G_{ij} > 0$: Las partículas de tipo $i$ son atraídas hacia las partículas de tipo $j$.
    \item $G_{ij} < 0$: Las partículas de tipo $i$ son repelidas por las partículas de tipo $j$.
    \item $G_{ij} = 0$: No hay interacción específica entre partículas de tipos $i$ y $j$ más allá de la repulsión a corta distancia.
\end{itemize}

La matriz no necesita ser simétrica, permitiendo a los estudiantes modelar relaciones asimétricas, como depredador-presa o parásito-huésped. En la implementación didáctica, los estudiantes pueden:

\begin{enumerate}
    \item Generar matrices aleatorias para observar comportamientos emergentes diversos.
    \item Diseñar matrices específicas para obtener comportamientos deseados.
    \item Analizar la relación entre la estructura de la matriz y los patrones resultantes.
\end{enumerate}

\section{Discretización para la simulación numérica}

La implementación computacional requiere discretizar las ecuaciones utilizando el método de Euler, introduciendo a los estudiantes en conceptos fundamentales de métodos numéricos:

\begin{equation}
\vec{r}_i(t + \Delta t) = \vec{r}_i(t) + \vec{v}_i(t) \Delta t
\end{equation}

\begin{equation}
\vec{v}_i(t + \Delta t) = \vec{v}_i(t) + \Delta t \left(\alpha\sum_{j \neq i} \vec{F}_{ij}(t) - \gamma\vec{v}_i(t)\right),
\end{equation}
donde $\Delta t$ es el paso de tiempo de la simulación.

Esta sección del proyecto permite a los estudiantes comprender:
\begin{itemize}
    \item La aproximación numérica de ecuaciones diferenciales
    \item Errores de truncamiento y estabilidad numérica
    \item Implementación de restricciones físicas como el límite de velocidad
    \item Condiciones de frontera y su implementación en código
\end{itemize}

\section{Pseudocódigo e implementación en Python}

\begin{algorithm}
\caption{Simulación de partículas con comportamiento emergente}
\label{alg:particles}
\begin{algorithmic}[1]
\State \textbf{Parámetros:} $N$ (número de partículas), $T$ (número de tipos), $L$ (tamaño del dominio), $r_{max}$ (radio máximo de interacción), $\alpha$ (intensidad de fuerza), $\beta$ (radio relativo de repulsión), $\gamma$ (amortiguamiento), $v_{max}$ (velocidad máxima), $\Delta t$ (paso de tiempo), $n_{frames}$ (número de frames)

\State \textbf{Inicializar:}
\State $\vec{r}_i \leftarrow$ posiciones aleatorias uniformes en $[0,L]^2$ para $i = 1,\ldots,N$
\State $\vec{v}_i \leftarrow$ velocidades aleatorias pequeñas para $i = 1,\ldots,N$
\State $\tau_i \leftarrow$ tipo aleatorio entre $0$ y $T-1$ para $i = 1,\ldots,N$
\State $G_{ij} \leftarrow$ valores aleatorios uniformes en $[-1,1]$ para $i,j = 0,\ldots,T-1$

\State \textbf{Para cada frame} $f = 1,\ldots,n_{frames}$:
\State \hspace{\algorithmicindent} \textbf{Para cada paso de tiempo dentro del frame:}
\State \hspace{\algorithmicindent}\hspace{\algorithmicindent} \textbf{Para cada partícula} $i = 1,\ldots,N$:
\State \hspace{\algorithmicindent}\hspace{\algorithmicindent}\hspace{\algorithmicindent} $\vec{F}_i \leftarrow \vec{0}$ \Comment{Inicializar fuerzas}
\State \hspace{\algorithmicindent}\hspace{\algorithmicindent}\hspace{\algorithmicindent} \textbf{Para cada partícula} $j = 1,\ldots,N$, $j \neq i$:
\State \hspace{\algorithmicindent}\hspace{\algorithmicindent}\hspace{\algorithmicindent}\hspace{\algorithmicindent} $\vec{d} \leftarrow \vec{r}_j - \vec{r}_i$ \Comment{Vector distancia}
\State \hspace{\algorithmicindent}\hspace{\algorithmicindent}\hspace{\algorithmicindent}\hspace{\algorithmicindent} \textbf{Ajustar $\vec{d}$ por condiciones periódicas:}
\State \hspace{\algorithmicindent}\hspace{\algorithmicindent}\hspace{\algorithmicindent}\hspace{\algorithmicindent} \textbf{Si} $d_x > L/2$ \textbf{entonces} $d_x \leftarrow d_x - L$
\State \hspace{\algorithmicindent}\hspace{\algorithmicindent}\hspace{\algorithmicindent}\hspace{\algorithmicindent} \textbf{Si} $d_x < -L/2$ \textbf{entonces} $d_x \leftarrow d_x + L$
\State \hspace{\algorithmicindent}\hspace{\algorithmicindent}\hspace{\algorithmicindent}\hspace{\algorithmicindent} \textbf{Si} $d_y > L/2$ \textbf{entonces} $d_y \leftarrow d_y - L$
\State \hspace{\algorithmicindent}\hspace{\algorithmicindent}\hspace{\algorithmicindent}\hspace{\algorithmicindent} \textbf{Si} $d_y < -L/2$ \textbf{entonces} $d_y \leftarrow d_y + L$

\State \hspace{\algorithmicindent}\hspace{\algorithmicindent}\hspace{\algorithmicindent}\hspace{\algorithmicindent} $r \leftarrow |\vec{d}|$ \Comment{Distancia}
\State \hspace{\algorithmicindent}\hspace{\algorithmicindent}\hspace{\algorithmicindent}\hspace{\algorithmicindent} \textbf{Si} $r > 0$ \textbf{y} $r < r_{max}$ \textbf{entonces}
\State \hspace{\algorithmicindent}\hspace{\algorithmicindent}\hspace{\algorithmicindent}\hspace{\algorithmicindent}\hspace{\algorithmicindent} \textbf{Si} $r < \beta \cdot r_{max}$ \textbf{entonces}
\State \hspace{\algorithmicindent}\hspace{\algorithmicindent}\hspace{\algorithmicindent}\hspace{\algorithmicindent}\hspace{\algorithmicindent}\hspace{\algorithmicindent} $F \leftarrow r/(\beta \cdot r_{max}) - 1$ \Comment{Repulsión a corta distancia}
\State \hspace{\algorithmicindent}\hspace{\algorithmicindent}\hspace{\algorithmicindent}\hspace{\algorithmicindent}\hspace{\algorithmicindent} \textbf{Sino}
\State \hspace{\algorithmicindent}\hspace{\algorithmicindent}\hspace{\algorithmicindent}\hspace{\algorithmicindent}\hspace{\algorithmicindent}\hspace{\algorithmicindent} $F \leftarrow G_{\tau_i, \tau_j} \cdot (1 - |2(r/r_{max} - 0.5)|)$ \Comment{Fuerza específica de tipo}
\State \hspace{\algorithmicindent}\hspace{\algorithmicindent}\hspace{\algorithmicindent}\hspace{\algorithmicindent}\hspace{\algorithmicindent} \textbf{Fin Si}
\State \hspace{\algorithmicindent}\hspace{\algorithmicindent}\hspace{\algorithmicindent}\hspace{\algorithmicindent}\hspace{\algorithmicindent} $\vec{F}_i \leftarrow \vec{F}_i + \alpha \cdot F \cdot \vec{d}/r$ \Comment{Acumular fuerza}
\State \hspace{\algorithmicindent}\hspace{\algorithmicindent}\hspace{\algorithmicindent}\hspace{\algorithmicindent} \textbf{Fin Si}
\State \hspace{\algorithmicindent}\hspace{\algorithmicindent}\hspace{\algorithmicindent} \textbf{Fin Para}

\State \hspace{\algorithmicindent}\hspace{\algorithmicindent} \textbf{Para cada partícula} $i = 1,\ldots,N$:
\State \hspace{\algorithmicindent}\hspace{\algorithmicindent}\hspace{\algorithmicindent} $\vec{v}_i \leftarrow (1-\gamma) \cdot \vec{v}_i + \vec{F}_i \cdot \Delta t$ \Comment{Actualizar velocidad con amortiguamiento}
\State \hspace{\algorithmicindent}\hspace{\algorithmicindent}\hspace{\algorithmicindent} \textbf{Si} $|\vec{v}_i| > v_{max}$ \textbf{entonces} $\vec{v}_i \leftarrow v_{max} \cdot \vec{v}_i / |\vec{v}_i|$ \Comment{Limitar velocidad}
\State \hspace{\algorithmicindent}\hspace{\algorithmicindent}\hspace{\algorithmicindent} $\vec{r}_i \leftarrow \vec{r}_i + \vec{v}_i \cdot \Delta t$ \Comment{Actualizar posición}
\State \hspace{\algorithmicindent}\hspace{\algorithmicindent}\hspace{\algorithmicindent} $\vec{r}_i \leftarrow \vec{r}_i \mod L$ \Comment{Aplicar condiciones periódicas}
\State \hspace{\algorithmicindent}\hspace{\algorithmicindent} \textbf{Fin Para}
\State \hspace{\algorithmicindent} \textbf{Fin Para cada paso de tiempo}
\State \hspace{\algorithmicindent} Guardar estado actual como frame de animación
\State \textbf{Fin Para cada frame}
\State Guardar animación en formato MP4
\end{algorithmic}
\end{algorithm}

A partir del pseudocódigo~\ref{alg:particles}, se puede implementar fácilmente el modelo de partículas en un lenguaje de programación. Hemos elegido Python por varias razones que lo convierten en una excelente opción para este tipo de proyectos. Python es un lenguaje multiplataforma, lo que significa que el código desarrollado puede ejecutarse sin problemas en diferentes sistemas operativos, lo que facilita su portabilidad y colaboración en equipos diversos. Además, Python es un lenguaje de alto nivel, lo que permite centrarse en la lógica del programa sin preocuparse por los detalles de la implementación a bajo nivel. 

Python también es multiparadigma, ya que admite tanto programación orientada a objetos como funcional, lo que permite una flexibilidad total en el estilo de programación según las necesidades del proyecto. La comunidad activa y extensa de Python es otro de sus grandes puntos fuertes; la disponibilidad de gran cantidad de librerías de código abierto, como \texttt{NumPy}, cuya implementación está altamente optimizada y vectorizada, facilita enormemente las operaciones numéricas y científicas, mejorando la eficiencia del código cuando es necesario.

Sin embargo, conscientemente hemos sacrificado la eficiencia computacional en algunos aspectos de la implementación, a fin de garantizar que el código sea lo más claro y comprensible posible para aquellos que estén aprendiendo o se inicien en este tipo de simulaciones. En varias partes del código, hemos optado por no vectorizar ciertas operaciones o utilizar técnicas avanzadas que podrían mejorar la eficiencia, para priorizar la comprensión y la transparencia en cada paso del algoritmo. En el apéndice del artíulo presentamos el código completo en Python.

\section{Implementación y desarrollo de competencias STEM}

La implementación de este modelo en un entorno educativo promueve el desarrollo de diversas competencias STEM:

\subsection{Competencias científicas}
\begin{itemize}
    \item Comprensión de principios físicos como fuerzas, movimiento y equilibrio
    \item Análisis de sistemas complejos y comportamientos emergentes
    \item Interpretación de resultados experimentales y simulaciones
    \item Formulación y prueba de hipótesis sobre cómo afectan los parámetros del sistema
\end{itemize}

\subsection{Competencias tecnológicas}
\begin{itemize}
    \item Programación en Python utilizando bibliotecas científicas (NumPy, Matplotlib)
    \item Uso de herramientas de visualización para representar datos complejos
    \item Desarrollo de algoritmos eficientes para sistemas de partículas
    \item Creación y gestión de animaciones científicas
\end{itemize}

\subsection{Competencias de ingeniería}
\begin{itemize}
    \item Diseño de sistemas basados en reglas para lograr comportamientos específicos
    \item Optimización de algoritmos para mejorar el rendimiento computacional
    \item Implementación de técnicas para estabilizar sistemas dinámicos
    \item Solución de problemas complejos mediante descomposición en partes más simples
\end{itemize}

\subsection{Competencias matemáticas}
\begin{itemize}
    \item Aplicación de ecuaciones diferenciales para modelar sistemas dinámicos
    \item Uso de álgebra lineal en la representación y manipulación de matrices
    \item Implementación de métodos numéricos para aproximación de soluciones
    \item Análisis cuantitativo de patrones emergentes y estructuras dinámicas
\end{itemize}

\section{Propuesta de actividades didácticas}

Para implementar este modelo como herramienta educativa, proponemos una secuencia de actividades didácticas:

\begin{enumerate}
    \item \textbf{Exploración conceptual:} Los estudiantes analizan los principios físicos y matemáticos del modelo antes de programarlo.
    
    \item \textbf{Desarrollo incremental:} Implementación paso a paso del algoritmo, comenzando con un sistema simple y añadiendo complejidad gradualmente.
    
    \item \textbf{Experimentación parametrizada:} Investigación sistemática de cómo diferentes parámetros afectan el comportamiento del sistema.
    
    \item \textbf{Diseño de matrices:} Creación de matrices de interacción específicas para generar comportamientos predefinidos, como agrupación, segregación o persecución.
    
    \item \textbf{Análisis de patrones:} Desarrollo de métodos cuantitativos para caracterizar los patrones emergentes.
    
    \item \textbf{Extensión del modelo:} Incorporación de nuevas características como evolución de parámetros, cambios ambientales o interacciones más complejas.
    
    \item \textbf{Proyecto final:} Aplicación del modelo a un fenómeno del mundo real, como comportamiento de bandadas, dinámica celular o fenómenos sociales.
\end{enumerate}

\section{Evaluación de competencias}

El proyecto permite evaluar el desarrollo de competencias STEM a través de:

\begin{itemize}
    \item \textbf{Código funcional:} Evaluación de la implementación correcta del algoritmo.
    
    \item \textbf{Documentación científica:} Informes que expliquen los principios subyacentes y los resultados observados.
    
    \item \textbf{Visualizaciones efectivas:} Calidad y claridad de las representaciones gráficas del sistema.
    
    \item \textbf{Experimentación:} Diseño y ejecución de experimentos sistemáticos para explorar el espacio de parámetros.
    
    \item \textbf{Extensión creativa:} Capacidad para modificar y expandir el modelo básico.
    
    \item \textbf{Comunicación:} Presentación clara de resultados y conclusiones.
\end{itemize}

\section{Conclusiones y aplicaciones educativas}

El modelo de partículas con comportamiento emergente representa una poderosa herramienta educativa para el desarrollo integrado de competencias STEM. Sus características clave lo hacen particularmente valioso:

\begin{itemize}
    \item \textbf{Interdisciplinariedad:} Integra conceptos de física, matemáticas, computación y biología.
    
    \item \textbf{Escalabilidad:} Puede adaptarse a diferentes niveles educativos, desde secundaria avanzada hasta educación superior.
    
    \item \textbf{Visualización:} Permite observar directamente fenómenos abstractos, facilitando su comprensión.
    
    \item \textbf{Experimentación:} Ofrece un entorno seguro para explorar hipótesis y probar ideas.
    
    \item \textbf{Relevancia:} Conecta con fenómenos del mundo real y tecnologías emergentes.
\end{itemize}

Este enfoque pedagógico favorece el aprendizaje activo, el pensamiento crítico y las habilidades de resolución de problemas, preparando a los estudiantes para abordar los complejos desafíos científicos y tecnológicos del siglo XXI. Además, el carácter lúdico y visual de estas simulaciones puede aumentar la motivación y el interés por las disciplinas STEM.

Futuras direcciones para este proyecto educativo incluyen el desarrollo de plataformas interactivas, la integración con tecnologías de realidad virtual para visualización avanzada, y la creación de comunidades de aprendizaje donde los estudiantes puedan compartir sus modelos y descubrimientos.

El modelo demuestra cómo comportamientos similares a los de sistemas vivos pueden emerger de reglas de interacción simples entre partículas, proporcionando un rico entorno de aprendizaje.

\newpage  
\appendix  
\section{Apéndice: Implementación en Python}
A continuación se muestra la implementación en Python del modelo de partículas basado en el pseudocódigo~\ref{alg:particles}.

\begin{lstlisting}[language=Python, caption={Implementación en Python del modelo de interacción de partículas.}, label={lst:algoritmo-genetico-continuo}]
import numpy as np
import matplotlib.pyplot as plt
from matplotlib.animation import FuncAnimation
from matplotlib import colors
import matplotlib.animation as animation

plt.style.use('dark_background')

# Parámetros de la simulación
num_particles = 200
box_size = 10.0
dt = 0.1
r_max = 1.0
alpha = 0.8  # Fuerza de interacción
beta = 0.3   # Radio de repulsión relativo a r_max
v_max = 0.1  # Velocidad máxima

# Número de tipos diferentes de partículas
num_types = 5

# Inicialización aleatoria de partículas
def initialize_particles(num_particles, box_size, num_types):
    particles = np.zeros((num_particles, 4))  # [x, y, vx, vy]
    
    # Posiciones aleatorias
    particles[:, 0] = np.random.uniform(0, box_size, num_particles)
    particles[:, 1] = np.random.uniform(0, box_size, num_particles)
    
    # Velocidades iniciales pequeñas
    particles[:, 2] = np.random.uniform(-0.1, 0.1, num_particles)
    particles[:, 3] = np.random.uniform(-0.1, 0.1, num_particles)
    
    # Asignar tipos aleatorios a las partículas (0 a num_types-1)
    particle_types = np.random.randint(0, num_types, num_particles)
    
    return particles, particle_types

# Matriz de interacción entre tipos de partículas
def generate_interaction_matrix(num_types):
    # Crear matriz de interacción aleatoria entre -1 y 1
    interaction_matrix = np.random.uniform(-1.0, 1.0, (num_types, num_types))
    return interaction_matrix

# Cálculo de fuerza entre partículas
def compute_force(r, type_i, type_j, interaction_matrix, r_max, beta):
    if r < beta * r_max:
        # Fuerza repulsiva a distancias cortas
        return r / (beta * r_max) - 1
    elif r < r_max:
        # Fuerza atractiva/repulsiva según la matriz de interacción
        return interaction_matrix[type_i, type_j] * (1 - abs(2 * (r / r_max - 0.5)))
    else:
        # Sin fuerza más allá de r_max
        return 0

# Actualizar posiciones y velocidades
def update_particles(particles, particle_types, interaction_matrix, box_size, dt, r_max, alpha, beta, v_max):
    num_particles = particles.shape[0]
    forces = np.zeros((num_particles, 2))
    
    # Calcular fuerzas entre todas las partículas
    for i in range(num_particles):
        for j in range(num_particles):
            if i != j:
                # Vector de distancia
                dx = particles[j, 0] - particles[i, 0]
                dy = particles[j, 1] - particles[i, 1]
                
                # Ajustar por condiciones de frontera periódicas
                if dx > box_size/2:
                    dx -= box_size
                elif dx < -box_size/2:
                    dx += box_size
                    
                if dy > box_size/2:
                    dy -= box_size
                elif dy < -box_size/2:
                    dy += box_size
                
                # Distancia entre partículas
                r = np.sqrt(dx**2 + dy**2)
                
                if r > 0 and r < r_max:
                    # Calcular fuerza
                    force_magnitude = compute_force(r, particle_types[i], particle_types[j], 
                                                   interaction_matrix, r_max, beta)
                    
                    # Componentes de la fuerza
                    forces[i, 0] += alpha * force_magnitude * dx / r
                    forces[i, 1] += alpha * force_magnitude * dy / r
    
    # Actualizar velocidades con amortiguamiento
    particles[:, 2] = (particles[:, 2] + forces[:, 0] * dt) * 0.8
    particles[:, 3] = (particles[:, 3] + forces[:, 1] * dt) * 0.8
    
    # Limitar velocidades
    speeds = np.sqrt(particles[:, 2]**2 + particles[:, 3]**2)
    for i in range(num_particles):
        if speeds[i] > v_max:
            particles[i, 2] = particles[i, 2] * v_max / speeds[i]
            particles[i, 3] = particles[i, 3] * v_max / speeds[i]
    
    # Actualizar posiciones
    particles[:, 0] += particles[:, 2] * dt
    particles[:, 1] += particles[:, 3] * dt
    
    # Aplicar condiciones de frontera periódicas
    particles[:, 0] = particles[:, 0] % box_size
    particles[:, 1] = particles[:, 1] % box_size
    
    return particles

# Función principal de simulación
def simulate_particle_life(num_frames=800):
    # Inicializar partículas y tipos
    particles, particle_types = initialize_particles(num_particles, box_size, num_types)
    
    # Generar matriz de interacción
    interaction_matrix = generate_interaction_matrix(num_types)
    
    # Configurar la figura para la animación
    fig, ax = plt.subplots(figsize=(8, 8))
    ax.set_xlim(0, box_size)
    ax.set_ylim(0, box_size)
    ax.set_title('Simulación de emergencia de vida')
    
    # Colores para diferentes tipos de partículas
    cmap = plt.cm.rainbow
    norm = colors.Normalize(vmin=0, vmax=num_types-1)
    
    # Crear scatter plot
    scatter = ax.scatter(particles[:, 0], particles[:, 1], 
                         c=particle_types, cmap=cmap, norm=norm, s=30)
    
    # Función de actualización para la animación
    def update(frame):
        nonlocal particles
        
        # Actualizar partículas 5 veces por frame para movimiento más suave
        for _ in range(5):
            particles = update_particles(particles, particle_types, interaction_matrix, 
                                        box_size, dt, r_max, alpha, beta, v_max)
        
        # Actualizar posición de las partículas en el gráfico
        scatter.set_offsets(particles[:, 0:2])
        
        # Mostrar el número de frame
        ax.set_title(f'Simulación de emergencia de vida - Frame: {frame}')
        
        return scatter,
    
    # Crear la animación
    ani = FuncAnimation(fig, update, frames=num_frames, blit=True, interval=50)
    
    # Guardar la animación como archivo MP4
    writer = animation.FFMpegWriter(fps=30, metadata=dict(artist='Me'), bitrate=1800)
    ani.save('particle_life_simulation.mp4', writer=writer)
    
    plt.close()
    
    print("Animación guardada como 'particle_life_simulation.mp4'")
    
    # Mostrar la matriz de interacción
    plt.figure(figsize=(6, 5))
    plt.imshow(interaction_matrix, cmap='coolwarm', vmin=-1, vmax=1)
    plt.colorbar(label='Fuerza de interacción')
    plt.title('Matriz de interacción entre tipos de partículas')
    plt.xlabel('Tipo de partícula')
    plt.ylabel('Tipo de partícula')
    plt.tight_layout()
    plt.savefig('interaction_matrix.png')
    plt.close()
    
    return interaction_matrix

# Ejecutar la simulación
if __name__ == "__main__":
    interaction_matrix = simulate_particle_life(num_frames=500)
    print("Simulación completada.")
\end{lstlisting}

A continuación se describen las principales partes del programa:

\subsection{Inicialización de partículas}

La función \texttt{initialize\_particles()} se encarga de generar las partículas de manera aleatoria en un espacio cuadrado de tamaño \texttt{box\_size}. Para cada partícula, se asignan las siguientes propiedades:
\begin{itemize}
    \item \textbf{Posición} (\(x, y\)): Generada aleatoriamente dentro de los límites de la caja.
    \item \textbf{Velocidad} (\(v_x, v_y\)): Inicializada con valores pequeños y aleatorios.
    \item \textbf{Tipo de partícula}: Un número aleatorio entre 0 y el número de tipos posibles, \texttt{num\_types}.
\end{itemize}

\subsection{Matriz de interacción}

La función \texttt{generate\_interaction\_matrix()} crea una matriz de interacción aleatoria entre los diferentes tipos de partículas. Los valores en la matriz están entre \([-1, 1]\), representando la intensidad de la fuerza de atracción o repulsión entre tipos.

\subsection{Cálculo de fuerzas}

La función \texttt{compute\_force()} calcula la fuerza que actúa entre dos partículas en función de la distancia \(r\) entre ellas. La fuerza se puede clasificar en tres casos:
\begin{itemize}
    \item \textbf{Fuerza repulsiva}: Si la distancia es menor a un umbral (\( \beta \cdot r_{\text{max}} \)).
    \item \textbf{Fuerza atractiva/repulsiva}: Depende de la matriz de interacción, si la distancia está dentro del rango \([ \beta \cdot r_{\text{max}}, r_{\text{max}} ]\).
    \item \textbf{Sin interacción}: Si la distancia supera el valor de \(r_{\text{max}}\).
\end{itemize}

\subsection{Actualización de partículas}

La función \texttt{update\_particles()} actualiza las posiciones y velocidades de las partículas en cada paso de la simulación. Se calcula la fuerza sobre cada partícula en función de las demás y se ajustan las velocidades y posiciones con una constante de amortiguamiento. Además, se aplica una condición de frontera periódica, de modo que las partículas que salen de la caja reaparecen en el lado opuesto.

\subsection{Animación de la simulación}

La función \texttt{simulate\_particle\_life()} orquesta la simulación y genera una animación de las partículas moviéndose en el espacio. Cada frame de la animación corresponde a un paso de la simulación, y las partículas se actualizan cinco veces por frame para un movimiento más fluido. La animación se guarda como un archivo \texttt{MP4} y la matriz de interacción entre partículas se guarda como una imagen \texttt{PNG}.

\subsection{Ejecución de la simulación}

La simulación se ejecuta en el bloque principal del programa, donde se llama a la función \texttt{simulate\_particle\_life()} para generar la animación y las visualizaciones de la matriz de interacción. La simulación se ejecuta con 500 frames por defecto.

\begin{lstlisting}[language=Python]
if __name__ == "__main__":
    interaction_matrix = simulate_particle_life(num_frames=500)
    print("Simulación completada.")
\end{lstlisting}

\begin{figure}[H] 
    \centering
    \includegraphics[width=0.5\textwidth]{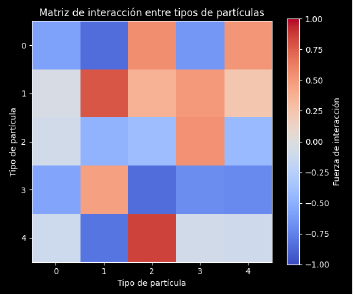} 
    \caption{Ejemplo de matriz de interacción.} 
    \label{fig:etiqueta_de_la_figura} 
\end{figure}

\begin{figure}[htbp]
  \centering
  \begin{subfigure}[b]{0.45\textwidth}
    \includegraphics[width=\textwidth]{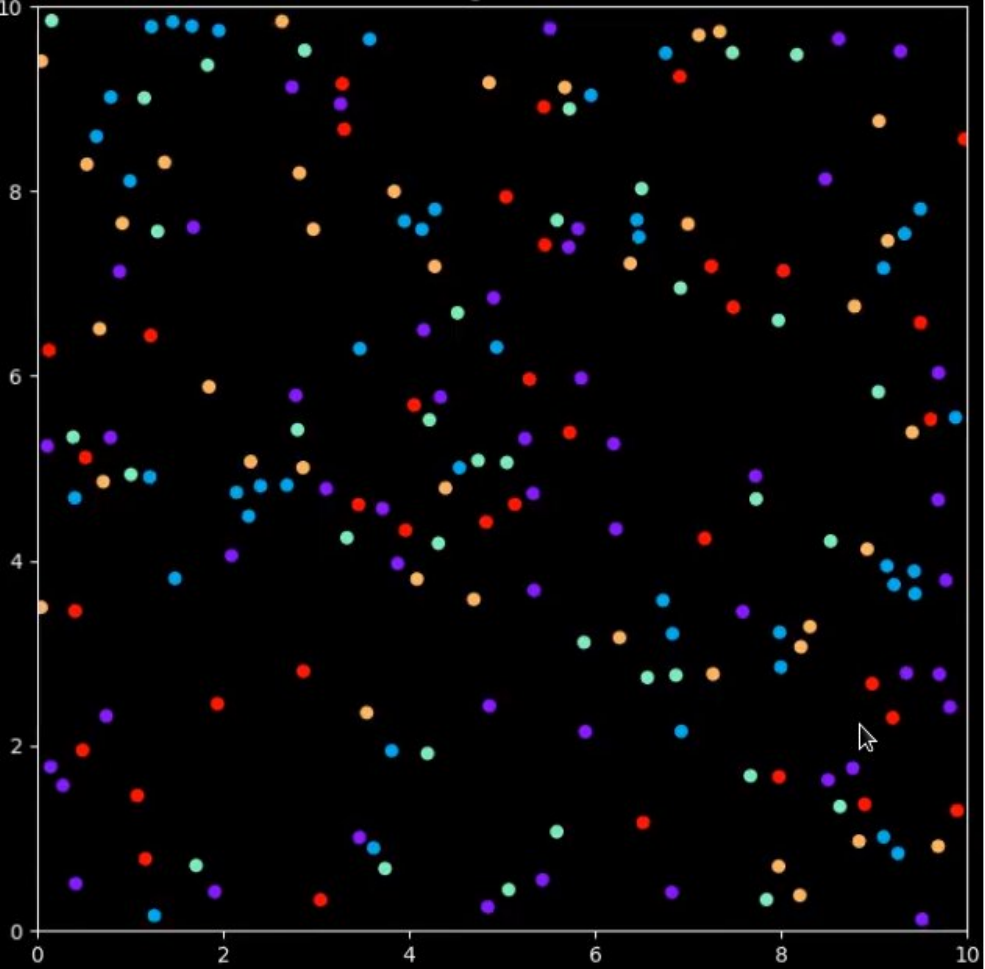}
    \caption{Configuración de las partículas en la iteración 0.}
    \label{fig:subfig1}
  \end{subfigure}
  \hfill 
  \begin{subfigure}[b]{0.45\textwidth}
    \includegraphics[width=\textwidth]{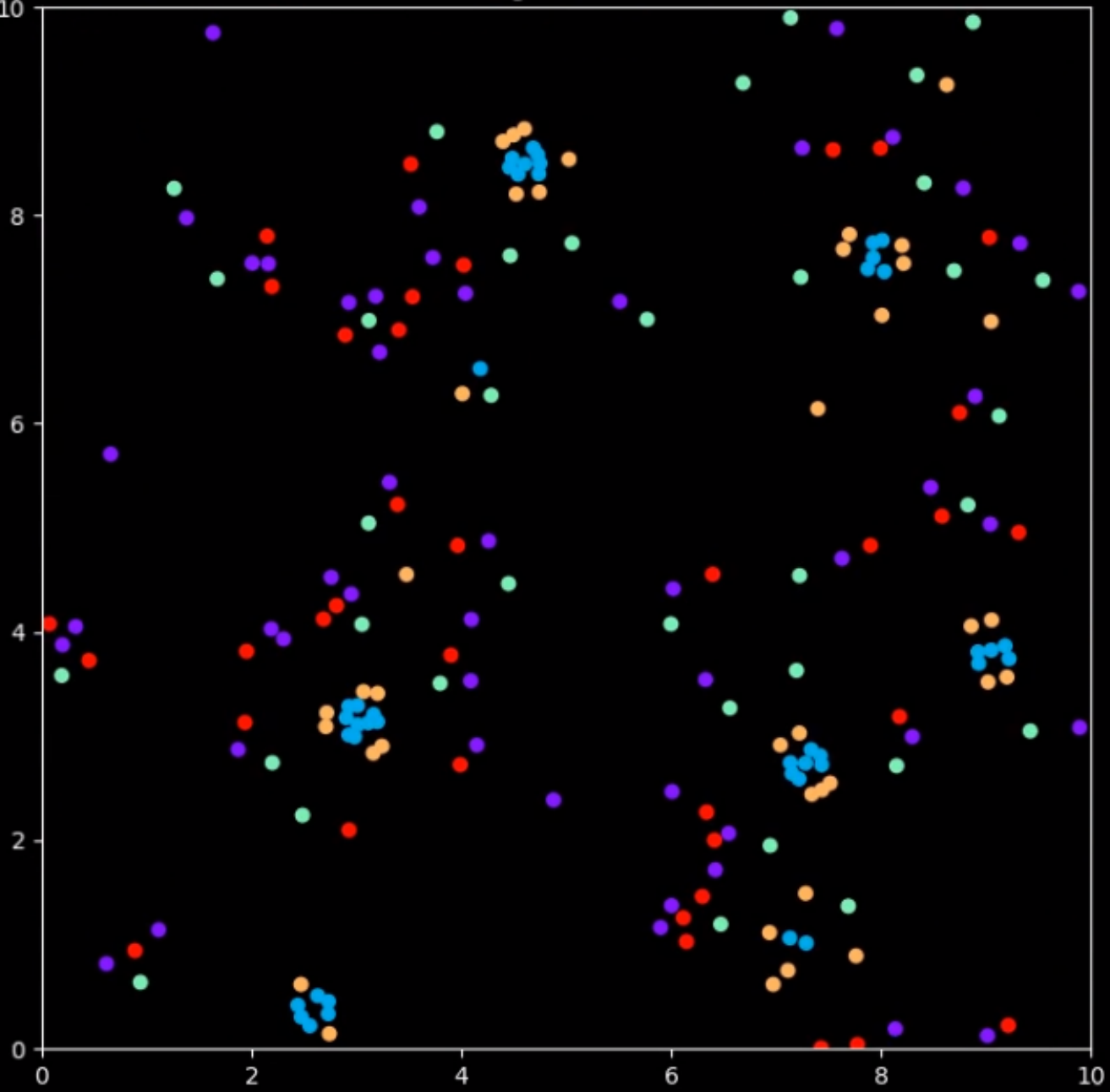}
    \caption{Configuración de las partículas en la iteración 100.}
    \label{fig:subfig2}
  \end{subfigure}
  \\[1ex] 
  \begin{subfigure}[b]{0.45\textwidth}
    \includegraphics[width=\textwidth]{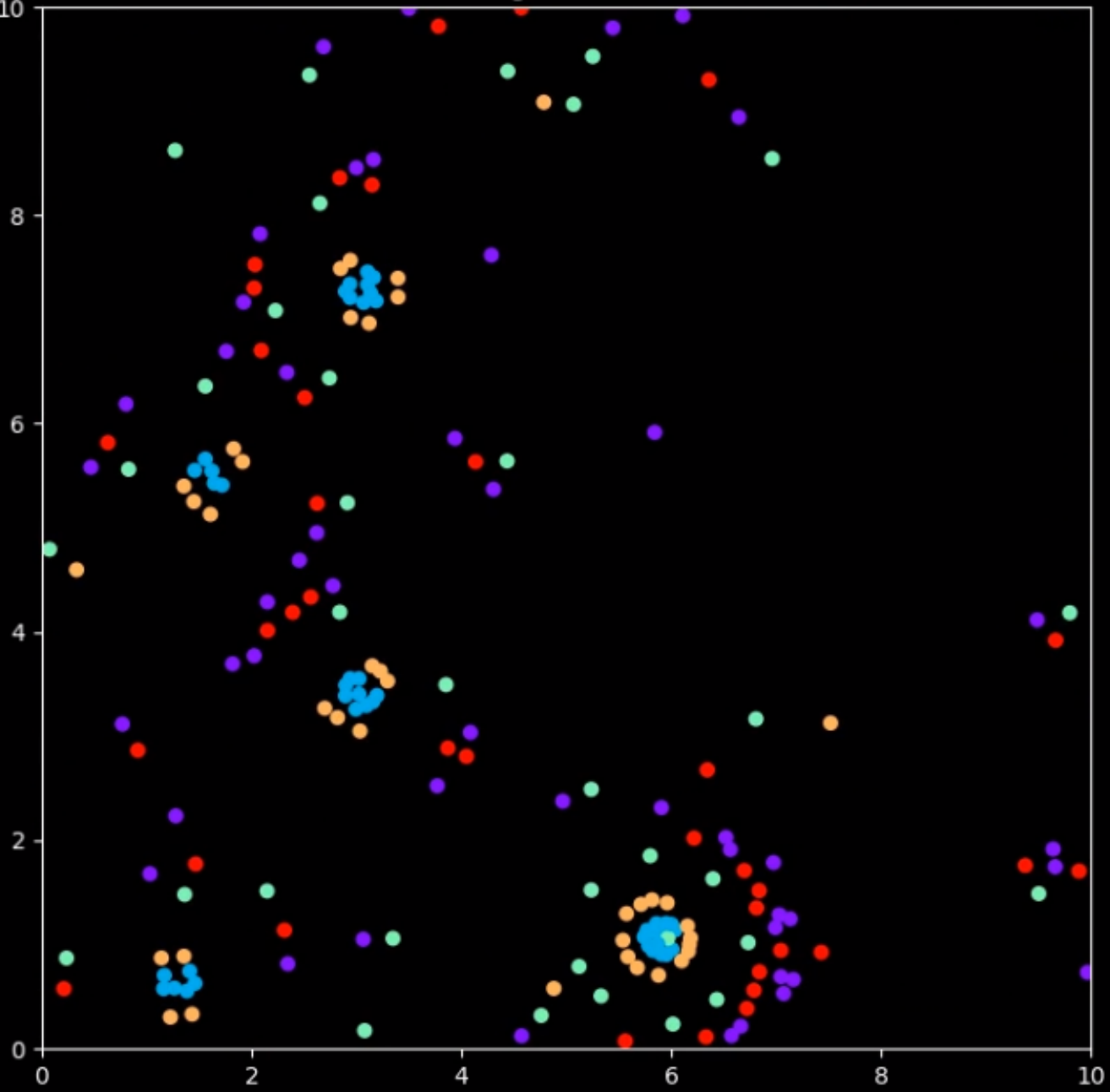}
    \caption{Configuración de las partículas en la iteración 200.}
    \label{fig:subfig3}
  \end{subfigure}
  \hfill 
  \begin{subfigure}[b]{0.45\textwidth}
    \includegraphics[width=\textwidth]{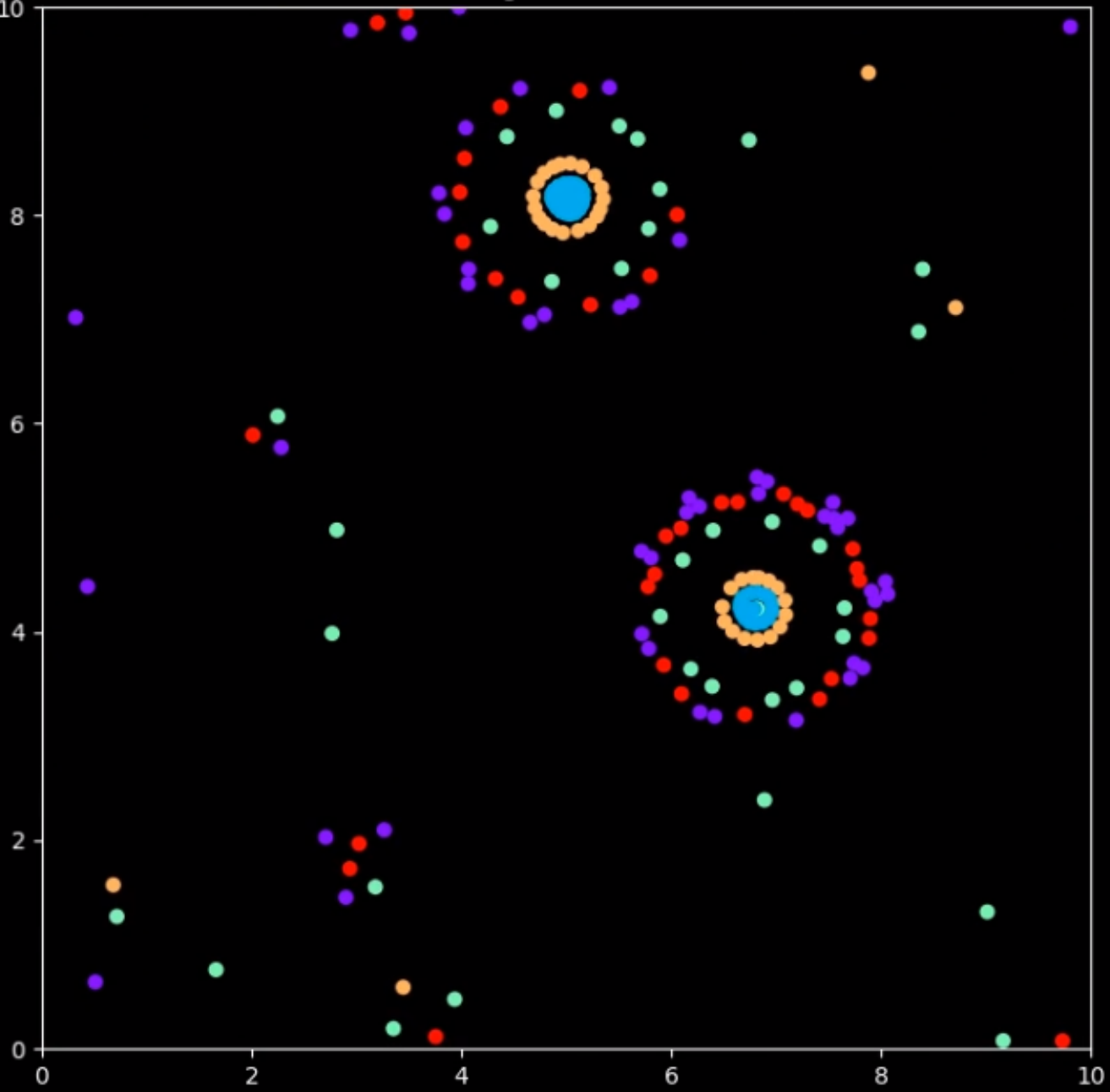}
    \caption{Configuración de las partículas en la iteración 300.}
    \label{fig:subfig4}
  \end{subfigure}
  \caption{Cuatro instancias del modelo después de ejecutar el código~\ref{lst:algoritmo-genetico-continuo}.}
  \label{fig:figura_principal}
\end{figure}

La figura \ref{fig:etiqueta_de_la_figura} muestra la matriz de interacción generada tras ejecutar el código correspondiente, el cual se encuentra detallado en el algoritmo \ref{lst:algoritmo-genetico-continuo}. Esta matriz refleja la relación de interacción entre los diferentes tipos de partículas, que se determina de manera aleatoria en función de los parámetros establecidos en la simulación.

Por otro lado, en la figura \ref{fig:figura_principal} se presenta la evolución del sistema dinámico a lo largo de cuatro iteraciones. En el primer paso, como se puede observar claramente, las partículas están distribuidas de manera aleatoria dentro del espacio definido. Sin embargo, a medida que avanza la simulación y las partículas interactúan entre sí, se puede apreciar la formación gradual de estructuras ordenadas. Este fenómeno es un claro ejemplo del comportamiento emergente en sistemas complejos, donde las interacciones locales entre las partículas conducen a la aparición de patrones globales y estructurados a lo largo del tiempo.

Ejecutando el código en reiteradas ocasiones, se pueden observar una gran variedad de escenarios, que podríamos conceptualizar como diferentes \textit{universos}. En cada ejecución, la configuración inicial aleatoria y la matriz de interacción cambian, lo que da lugar a un comportamiento completamente distinto. En algunos de estos escenarios, las partículas tienden a formar estructuras muy complejas y organizadas, mientras que en otros, las partículas permanecen dispersas, formando configuraciones más simples y distantes. 

Además, el código permite modificar varios parámetros de la simulación. Por ejemplo, aumentando el número de partículas que interactúan, se puede observar cómo la complejidad de las estructuras generadas crece considerablemente. De igual forma, ajustando los valores de los parámetros de interacción, como la fuerza de atracción y repulsión, o modificando el tamaño del espacio de simulación, se pueden obtener nuevos comportamientos que varían desde patrones altamente organizados hasta configuraciones casi caóticas. Estos experimentos permiten explorar un amplio espectro de posibles realidades, demostrando la rica diversidad de comportamientos que emergen de simples reglas locales de interacción.

\end{document}